\documentclass[12pt]{article}

\usepackage{amsfonts,bm}
\usepackage{amscd}
\usepackage{amsmath}

\usepackage{framed}
\usepackage{color}
\usepackage{slashed}
\usepackage{pdflscape}
\usepackage{enumerate}
\usepackage{multirow}
\usepackage[backref,pagebackref,linkcolor=blue]{hyperref}
\usepackage{mathrsfs}
\usepackage{comment}

\def\sss{\scriptscriptstyle}

\def\wt#1{\widetilde{#1}}

\def\sss{\scriptscriptstyle}

\def\be{\begin{equation}}
\def\ee{\end{equation}}
\def\beq{\begin{equation}}
\def\eeq{\end{equation}}
\def\bea{\begin{eqnarray}}
\def\eea{\end{eqnarray}} 
\def\beqa{\begin{equation}\begin{array}{l}}
\def\eeqa{\end{array}\end{equation}}

\def\eqn#1{(\ref{#1})}

\def\eqref#1{eq.~(\ref{eq:#1})}

\def\G{{\it\Gamma}}

\def\nn{\nonumber}

\newcommand{\Rho}{{\mbox{\sf P}}}

\def\sideremark#1{\ifvmode\leavevmode\fi\vadjust{\vbox to0pt{\vss
 \hbox to 0pt{\hskip\hsize\hskip1em
 \vbox{\hsize3cm\tiny\raggedright\pretolerance10000
 \noindent #1\hfill}\hss}\vbox to8pt{\vfil}\vss}}}%
                        
\begin{document}

\thispagestyle{empty}
\begin{flushright}
\framebox[.9\width]{\tiny \tt \begin{tabular}{c}CALT  68-2884\\ BRX-TH-661\end{tabular} }
\end{flushright}

\vspace{.8cm}
\setcounter{footnote}{0}

\begin{center}
{\Large{\bf Gravitational- and Self- Coupling  of  Partially
Massless~Spin~2
 }
    }\\[10mm]

{\sc S. Deser$^\sharp$, E. Joung$^\flat$
and A. Waldron$^\natural$
\\[6mm]}

{\em\small  
$^\sharp$Lauritsen Lab,
Caltech,
Pasadena CA 91125
 and Physics Department, Brandeis University, Waltham,
MA 02454, 
USA\\ \href{mailto:deser@brandeis.edu}{\tt deser@brandeis.edu}}\\[5mm]
{\em\small  
$^\flat$ Scuola Normale Superiore and INFN,
Piazza dei Cavalieri 7, 56126 Pisa, Italy \\
\href{mailto:euihun.joung@sns.it}{\tt euihun.joung@sns.it}}\\[5mm]
{\em\small  
$^\natural$Department of Mathematics,
University of California, Davis, CA 95616,
USA\\ \href{mailto:wally@math.ucdavis.edu}{\tt wally@math.ucdavis.edu}}\\[5mm]

\bigskip

\bigskip

{\sc Abstract}\\
\end{center}

{\small
\begin{quote}

We show that higher spin systems specific to cosmological spaces are subject to the same problems as models with Poincar\'e limits. In particular, we analyse partially massless (PM) spin~2  and find that both its gravitational coupling and nonlinear extensions suffer from the usual [background- and self-coupling] difficulties: Consistent free field propagation does not extend beyond  background Einstein geometries.
Then, using conformal Weyl gravity (CG),  which consists of relative ghost~PM  and  graviton excitations, we find that avoiding graviton-ghosts 
restricts
 CG-generated~PM self-couplings to the usual, safe, Noether current cubic ones.

\bigskip

{\tt PACS: 04.62.+v, 04.50.-h, 04.50.Kd, 04.62.+v, 02.40.-k}

\bigskip

\end{quote}
}

\newpage





\section{Introduction}

The consistency difficulties of massless and massive higher spin fields in $d=4$ are by now well-explored, both regarding their coupling to gravity and other fields as well  as possible self-interactions. Our aim here is to investigate these problems for
partially massless~(PM) theories~\cite{Deser:1983tm,Deser:2001pe}, which have the novel feature that their (anti) de Sitter~((A)dS) higher spin representations have no
direct Poincar\'e counterparts.
 For this we employ Weyl---conformal---gravity (CG) as a tool.  Even though~CG is  physically unacceptable 
  (being fourth derivative order, its physical excitations are relatively ghost-like) 
   it can be safely used when one of its two, graviton and~PM~\cite{Maldacena:2011mk}, components can be fixed, while studying the other~\cite{us}. 

We will begin by reviewing~PM  and then show that  it precisely characterizes~CG solutions that are not conformally Einstein spaces. We then explain,  using recent mathematical tools, how~CG can be safely exploited for our consistency analyses of~PM. The first question: \begin{quote}{\it what are the most general geometrical fixed backgrounds in which~PM consistently propagates?}\end{quote} can then be answered--they are essentially restricted to Einstein spaces. The second consistency question: \begin{quote}{\it can one define a self-interacting version of the free field, even in Einstein vacuum?}\end{quote} will then be addressed,
yielding a  minor triumph as well: only the usual cubic, abelian Noether current-field coupling is generated via~CG. We conclude with  speculations regarding~PM's possible cosmological and formal uses.

\section{Review of~PM and its~CG embedding}

The~PM tensor field $\varphi_{\mu\nu}$ dynamics are defined in any Einstein background by the action 
\begin{eqnarray}\label{PMEQ}&=-\int\sqrt{-g}\Big[& \:
\tfrac12\big(\nabla_\rho \varphi_{\mu\nu}\big)^2-\big(\nabla^\nu \varphi_{\mu\nu}\big)^2
+\nabla^\mu\varphi \, \nabla^\nu \varphi_{\mu\nu}-\tfrac12\big(\nabla_\mu \varphi\big)^2\\
&&\!+\varphi^{\mu\nu}W_{\rho\mu\nu\sigma}\varphi^{\rho\sigma}+\tfrac{2\Lambda}3\big[\big(\varphi_{\mu\nu}\big)^2-\tfrac14\varphi^2\big]
\Big]\, ,\quad \varphi:=\varphi^\rho_\rho\, ,
\end{eqnarray}
and field equations  
\begin{eqnarray*}
\Delta\,\varphi_{\mu\nu}
- 2\,\nabla_{(\mu}\,\nabla^\rho\,\varphi_{\nu)\rho}+ g_{\mu\nu}\,\nabla^\rho\,\nabla^\sigma\, \varphi_{\rho\sigma}
+\nabla_\mu\,\nabla_\nu\,\varphi 
-g_{\mu\nu}\,\Delta\,\varphi\\[2mm]
-\,2\,W_{\rho\mu\nu\sigma}\,\varphi^{\rho\sigma}
-\tfrac43\,{\Lambda}\,(\varphi_{\mu\nu} -\tfrac14\,g_{\mu\nu}\varphi)=0\, ,
\end{eqnarray*}  
where $W_{\mu\nu\rho\sigma}$ is the Weyl tensor.
This system is invariant under a double derivative gauge transformation
 \begin{equation}\label{pm_gauge}
 \delta \varphi_{\mu\nu}=
 \big(\nabla_\mu\partial_\nu + \tfrac\Lambda 3\,g_{\mu\nu}\big)\alpha(x)\, ,
 \end{equation}  
  which is the tuned sum of a metric fluctuation diffeomorphism (with parameter $\partial_\mu \alpha(x)$) and a conformal  transformation.
  This system is a hybridization of strictly massless and normal massive, Fierz-Pauli,
spin 2.
Indeed,  there are three varieties of spin 2 excitations in~dS: massive, massless and~PM~\cite{Deser:1983tm,Deser:2001pe}. 
 In~dS,~PM propagates
 lightlike, positive energy (inside the maximally accessible intrinsic horizon), helicity~$\pm 2,\pm1$ excitations in a unitary representation of the isometry group~\cite{Higuchi:1986py,Deser:2001xr,Deser:2001wx,Deser:2003gw}. 
  This degree of freedom (DoF) count relies on the  gauge invariance~\eqn{pm_gauge} and the divergence constraint $\nabla^\mu \varphi_{\mu\nu}=\nabla_\nu \varphi$ implied by integrability of~\eqn{PMEQ}.
 
 Interactions of~PM in four dimensions are particularly interesting because it is rigidly $SO(4,2)$ conformally invariant~\cite{Deser:2004ji}, just like its vector Maxwell counterpart. In fact,~PM can be
coupled to charged matter fields~\cite{Deser:2006zx} (see also~\cite{Zinoviev:2009hu}). [Forming  non-abelian multiplets is still an open problem.]
Instead, we will be concerned with its  self interactions, whose  cubic vertices  
 were first given in~\cite{Zinoviev:2006im} using a St\"uckelberg approach.\footnote{A general calculus of higher derivative~PM cubic vertices was developed in~\cite{Joung:2012rv}. Also, it has recently been suggested that a~PM limit of  putative massive gravity theories could be a candidate for an interacting~PM theory~\cite{deRham:2012kf}.}

 Since Weyl transformations underlie~PM's invariances (see~\eqn{pm_gauge}),~CG is a natural tool
 for studying its interactions.
While~CG always has six excitations, the detailed spectra are background-dependent.
About flat space, it has two massless tensors and a photon with the same signature as one of them~\cite{2+2+2}, while in constant curvature backgrounds there is still a (cosmological) graviton, but now the (tensor+photon) combination becomes the~PM mode with helicities $(\pm2,\pm1)$. In each case, the two sets of  modes are relatively ghost-like. The relative sign between~PM's helicities depends on that of~$\Lambda$: In AdS, one can truncate the solution space to just
the unitary, massless graviton~\cite{Maldacena:2011mk,Lu:2011ks,Metsaev:2007fq} (for related analysis of higher derivative theories see~\cite{Lu:2011zk,Hyun:2011ej}).
The dS story is the more interesting one because we can truncate, leaving either mode unitary; keeping the unitary PM mode, is the relevant case here.

 The~CG action is 
\begin{eqnarray}\label{W2action}
S[g] = \tfrac18\int  \sqrt{-g}\ W^{\mu\nu\rho\sigma}\,W_{\mu\nu\rho\sigma} 
\, =\,  \tfrac14\int \sqrt{-g}\,\left( R^{\mu\nu}\,R_{\mu\nu}-{\tfrac13}\,R^2 \right),
\end{eqnarray}
and its field equation is the vanishing of the Bach tensor,
\begin{equation*}\label{Bach}
B_{\mu\nu}:=-\Delta\,\Rho_{\mu\nu}+
\nabla^\rho\,\nabla_{(\mu}\,\Rho_{\nu)\rho}+W_{\rho\mu\nu\sigma}\,\Rho^{\rho\sigma}\,,\qquad \Rho_{\mu\nu}:=\frac12 (R_{\mu\nu}-\frac16\, g_{\mu\nu} R)\, .
\end{equation*}
The Schouten tensor $\Rho_{\mu\nu}$ measures the difference between Riemann and Weyl tensors, $R_{\mu\nu\rho\sigma}-W_{\mu\nu\rho\sigma}=g_{\mu\rho}\Rho_{\nu\sigma}-g_{\nu\rho}\Rho_{\mu\sigma}+g_{\nu\sigma}\Rho_{\mu\rho}-g_{\mu\sigma}\Rho_{\nu\rho}$, and is a mainstay of conformal models in all dimensions: its variation is a pure (double) gradient, 
\begin{equation*}\delta g_{\mu\nu} = 2\,\alpha\,g_{\mu\nu}
\label{rhoweyl}\quad
\Rightarrow\quad
\delta \Rho_{\mu\nu}=-\nabla_\mu\,\partial_\nu\,\alpha\, .
\end{equation*}
The Bach tensor $B _{\mu\nu}$ is, of course, invariant under this rescaling. For our purposes, it is more convenient to work with the cosmological Schouten tensor,
\begin{equation}\label{varphi}
\varphi_{\mu\nu}:= -\Rho_{\mu\nu}+\tfrac \Lambda 6\,g_{\mu\nu}\, ,
\end{equation}
in terms of which $B _{\mu\nu}$ reads
\begin{eqnarray}\label{pm_eom}
B_{\mu\nu}(g,\varphi)
&=&\Delta\,\varphi_{\mu\nu}
- 2\,\nabla_{(\mu}\,\nabla^\rho\,\varphi_{\nu)\rho}+ g_{\mu\nu}\,\nabla^\rho\,\nabla^\sigma\, \varphi_{\rho\sigma}
+\nabla_\mu\,\nabla_\nu\,\varphi 
-g_{\mu\nu}\,\Delta\,\varphi\nonumber\\[2mm]
&&-\,2\,W_{\rho\mu\nu\sigma}\,\varphi^{\rho\sigma}
-\tfrac43\,{\Lambda}\,(\varphi_{\mu\nu} -\tfrac14\,g_{\mu\nu}\varphi) \, + \, O\big(\varphi^2\big)
\, . \label{BachPM}
\end{eqnarray}
Consider now configurations  such that the metric is close to an Einstein one with cosmological constant $\Lambda$ (it is important to note that the set of Bach flat, but non-Einstein metrics is non-empty, see~\cite{Nurowski:2000cq}). Then, by~\eqn{varphi}, $\varphi_{\mu\nu}$ is a small excitation and its field equation~\eqn{pm_eom} is precisely 
the PM one in this background, the Schouten tensor's Weyl transformation implying the PM gauge invariance~\eqn{pm_gauge}.
We have now recovered~CG's (linearized)~PM subsector by holding the metric constant (or in other words, setting the metric to a non-dynamical background field). This
 key fact  motivates our use of~CG as a probe of~PM for two basic higher spin questions:
How general are the geometries in which it can propagate consistently?
Does~CG provide a useful starting-point for studying possible self-interactions of~PM?

We  next answer the first question: we  use~CG to generate a list of increasingly general metrics, from dS to Einstein to Bach, and show that there is indeed a natural barrier--one that is much closer to Einstein than, as one might reasonably conjecture, to Bach.

\section{PM in a background}

It has been established that there exist Weyl invariant field equations enjoying a double derivative gauge invariance in Bach-flat backgrounds~\cite{Gover:2006fa}. This result suggests that Bach-flat is the most general background supporting consistent (linear)~PM  propagation. In detail, the operator from scalars to trace-free symmetric tensors,
$$
{\bm P}_{\mu\nu}:= \nabla_{\{\mu}\,\partial_{\nu\}} + 
\Rho_{\{\mu\nu\}}\,,
$$
permits a factorization of the Bach tensor as
\begin{eqnarray}\label{detour_complex}
B_{\mu\nu}&\ =&\ {\bm M}_{\mu\nu}^{\rho\sigma} \,{\bm P}_{\rho\sigma}\, ,\\[2mm]
{\bm M}^{\rho\sigma}_{\mu\nu}&:=&\ \delta^\rho_{\{\mu}\delta^{\sigma\phantom{\rho}}_{\nu\}}\Delta
-\delta_{\{\mu}^\rho \nabla^\sigma\, \nabla_{\nu\}}^{\phantom{\rho}}
-\tfrac13\, \delta_{\{\mu}^\rho \nabla_{\nu\}}^{\phantom{\rho}}\nabla^\sigma  -W^\rho{}_{\mu\nu}{}^\sigma\, .\nn
\end{eqnarray}
We observe that~${\bm M}_{\mu\nu}^{\rho\sigma}$ gives the non-linear answer to the  question posed in the Introduction: characterizing Bach-flat metrics
that are not conformally Einstein (the latter are characterized in~\cite{Kozameh:1985jv} and~\cite{BEG}). We see that those require the range of~${\bm P}_{\mu\nu}$ to intersect the kernel of~${\bm M}_{\mu\nu}$; the operator~${\bm M}_{\mu\nu}$ is also conformally invariant and maps trace-free symmetric tensors to trace-free symmetric tensors.
Physically, it implies that the field equation
\begin{equation}\label{ym_cmplx_eom}
{\bm M}^{\rho\sigma}_{\mu\nu}\,  \wt\varphi_{\rho\sigma} = \Delta\, \wt \varphi_{\mu\nu}-\nabla^\sigma\, \nabla_{\{\mu}\,\wt\varphi_{\nu\}\sigma} 
-\tfrac 13\, \nabla_{\{\mu}\,\nabla^\sigma\, \wt\varphi_{\nu\}\sigma}
-W^\rho{}_{\mu\nu}{}^\sigma\,\wt\varphi_{\rho\sigma}=0\, ,
\end{equation}
for a trace-free symmetric tensor~$\wt \varphi_{\mu\nu}=:\varphi_{\{\mu\nu\}}=\varphi_{\mu\nu}-\frac14 g_{\mu\nu} \varphi$, enjoys the double derivative gauge invariance (and associated double derivative Bianchi identity)
$$
\delta \wt\varphi_{\mu\nu} = {\bm P}_{\mu\nu} \alpha = \big(\nabla_{\{\mu}\nabla_{\nu\}} +\wt \Rho_{\mu\nu}\big) \alpha\, ,
$$
in Bach-flat backgrounds. This was the  motivation for  our original conjecture that~PM fields could propagate in them. We now proceed to  disprove it and give necessary consistency conditions for~PM-compatible backgrounds.

The Bach tensor, since it arises from a metric variational principle, is necessarily divergence-free, $\nabla^\mu {\bm M}_{\mu\nu}^{\rho\sigma}\, {\bm P}_{\rho\sigma}=0$. However, it is neither true that~$\nabla^\mu {\bm M}_{\mu\nu}^{\rho\sigma}=0$, nor even that~$\nabla^\mu {\bm M}_{\mu\nu}^{\rho\sigma}=O(\nabla)$ (rather this operator is cubic in derivatives). But consistent~PM propagation relies on a divergence constraint\footnote{The DoF count for~PM starts with  ten off-shell fields~$\varphi_{\mu\nu}$, minus four DoF thanks to the divergence  constraint~$\nabla^\mu \varphi_{\mu\nu}=\nabla_\nu \varphi$,
minus two further DoF by the scalar gauge invariance, yielding a total of four on-shell excitations.}; for 
 a~PM field equation (derived from an action) this requirement is precisely expressed by the condition~$\nabla^\mu {\bm M}_{\mu\nu}^{\rho\sigma}=O(\nabla)$.
 
The failure of the field equation~\eqn{ym_cmplx_eom} to imply an appropriate divergence constraint does not yet rule out~PM
fields interacting with backgrounds more general than Einstein spaces, because we may still enlarge the space of field equation  and
gauge operators,~${\bm M}_{\mu\nu}^{\rho\sigma}$ and~${\bm P}_{\mu\nu}$ respectively, by relaxing their trace-free and conformal invariance properties.
To test this, we make the following generalization
\begin{eqnarray*}
{\bm M}^\prime{}_{\mu\nu}^{\rho\sigma}\!\!\!&=&{\bm G}_{\mu\nu}^{\rho\sigma}
-\big(\delta^\rho_{(\mu}\,\delta^{\sigma\phantom{\rho}}_{\nu)}-g_{\mu\nu}\,g^{\rho\sigma}\big)\, \Rho
+\alpha_1\,\delta^\rho_{(\mu}\,\wt\Rho{}_{\nu)}^{\sigma\phantom{\rho}} 
+\alpha_2\,\big(g_{\mu\nu}\,\wt\Rho{}^{\rho\sigma}+\wt\Rho_{\mu\nu}\,g^{\rho\sigma}\big)
\ ,\\[2mm]
{\bm P}^\prime_{\mu\nu}&=&\nabla_\mu\,\partial_\nu + \tfrac12\, \Rho\,  g_{\mu\nu}  + \beta\,\wt\Rho_{\mu\nu}\, ,
\end{eqnarray*}
where the cosmological Einstein operator 
\begin{eqnarray}
{\bm G}_{\mu\nu}^{\rho\sigma}&:=&\big(\delta^\rho_{(\mu}\delta^{\sigma\phantom{\rho}}_{\nu)}-g_{\mu\nu}g^{\rho\sigma}\big)\big(\Delta-\Rho\big)
-2\,\nabla_{(\mu} \nabla^\rho \delta^\sigma_{\nu)}+\nabla_{(\mu}\nabla_{\nu)} g^{\rho\sigma}+g_{\mu\nu}\nabla^\rho\nabla^\sigma
\nonumber\\[1mm]&&-\,2\,W^\rho{}_{\mu\nu}{}^\sigma
-8\,\widetilde\Rho{}_{\{\mu}^\rho\delta^{\sigma\phantom{\rho}}_{\nu\}}-\tfrac32\,  g_{\mu\nu} \Rho\,  g^{\rho\sigma}\, ,\label{Gop}
 \end{eqnarray}
 is identically conserved,
~$$
 \nabla^\mu {\bm G}_{\mu\nu}^{\rho\sigma}=0\, ,
~$$
in Einstein backgrounds. The equation of motion of cosmological Einstein gravity linearized about an Einstein metric 
is~${\bm G}_{\mu\nu}^{\rho\sigma} \,\varphi_{\rho\sigma}=0$.

The above ansatz is the most general one obeying the following requirements:
\begin{enumerate}
\item The operators ~${\bm M}^\prime{}_{\mu\nu}^{\rho\sigma}$ and~${\bm P}^\prime_{\mu\nu}$ are second order in~$\nabla$ or derivatives on the
metric~$g_{\mu\nu}$.
\item The operator~${\bm M}^\prime{}_{\mu\nu}^{\rho\sigma}$ is self-adjoint,  to ensure the existence of an action principle.
\item The divergence~$\nabla^\mu {\bm M}^\prime{}_{\mu\nu}^{\rho\sigma}$ is an operator no more than linear in~$\nabla$ so
that solutions of~${\bm M}^\prime{}_{\mu\nu}^{\rho\sigma}\,\varphi_{\mu\nu}=0$ obey a first order constraint.
\item The operator product~${\bm M}^\prime{}_{\mu\nu}^{\rho\sigma}{\bm P}^\prime_{\rho\sigma}$ vanishes when~$g_{\mu\nu}$ is
an Einstein metric; this fixes their leading terms to be operators corresponding to the linear~PM equation of motion~\eqn{pm_eom}
and its double derivative gauge invariance~\eqn{pm_gauge}. The remaining freedom in the ansatz therefore depends  only on the trace-free
Schouten tensor~$\wt \Rho_{\mu\nu}$, since that quantity vanishes for Einstein metrics.
\end{enumerate} 

It remains to compute the product 
${\bm M}^\prime{}_{\mu\nu}^{\rho\sigma}\,{\bm P}^\prime_{\rho\sigma}$. The result can be arranged as an expansion in the gradient 
operator~$\nabla$. By construction, terms of order~$\nabla^4$ and~$\nabla^3$ necessarily vanish. Prefactors of the terms order~$\nabla^2$ 
only involve~$\wt \Rho_{\mu\nu}$ which we are now assuming to be non-vanishing, since we wish to investigate metrics that are {\it not} Einstein:
we must choose the constants~$(\alpha_1,\alpha_2,\beta)$ accordingly and find
$$
\alpha_1=4+2\beta\, , \qquad \alpha_2=-\beta\, .
$$
The analysis of terms order~$\nabla$ and lower is more complicated. First we consider the trace~$g^{\mu\nu}{\bm M}^\prime{}_{\mu\nu}^{\rho\sigma}{\bm P}^\prime_{\rho\sigma}$ at order~$\nabla$ and find~$3\beta (\nabla_\rho \Rho)\nabla^\rho$. There are two possibilities, either~$\beta=0$ or the background metric has constant scalar curvature.
Since the latter  would rule out the~PM conjecture in question,  we choose~$\beta=0$.
We then find~$g^{\mu\nu}{\bm M}^\prime{}_{\mu\nu}^{\rho\sigma}{\bm P}^\prime_{\rho\sigma}=-3(\Delta \Rho)$, which requires the scalar curvature to be harmonic,
and hence also rules out the conjecture.

Having excluded Bach-flat backgrounds,
we may still investigate whether some  condition stronger  than Bach-flat, but still less stringent than  Einstein, could yield consistent propagation.
The terms remaining at order~$\nabla$ in~${\bm M}^\prime{}_{\mu\nu}^{\rho\sigma}{\bm P}^\prime_{\rho\sigma}$ are
$$
\beta g_{\mu\nu} (\nabla_\rho \Rho)\nabla^\rho-(\beta-2)(\nabla_{(\mu}\Rho)\nabla_{\nu)}
+2(\beta-1)(\nabla_\rho \Rho_{\mu\nu})\nabla^\rho-2\beta(\nabla_{(\mu}\Rho_{\nu)\rho})\nabla^\rho\, ;
$$
clearly no choice of~$\beta$ removes all of them. Instead, we can restrict the background, one option being to Ricci-symmetric spaces, defined by~$\nabla_\rho \Rho_{\mu\nu}$ $=$~$0$. This condition is weaker than  Einstein, but need not imply Bach-flat.
However, even then we must cancel all terms in~${\bm M}^\prime{}_{\mu\nu}^{\rho\sigma}{\bm P}^\prime_{\rho\sigma}$ of order~$\nabla^0$. In {\it general} backgrounds these are
\begin{eqnarray*}
-\beta\,B_{\mu\nu}+2\beta^2\,\Rho^\rho_{(\mu}\Rho^{\phantom{\rho}}_{\nu)\rho}-\tfrac12 (\beta-1)(\beta+3)\, \Rho \, \Rho_{\mu\nu}
-\tfrac12 (\beta-2)\, \nabla_\mu\partial_\nu \Rho\\
 +\, g_{\mu\nu}\big[\tfrac12(\beta-2)\,\Delta \Rho-\beta(\beta+1)\,\Rho_{\rho\sigma}\Rho^{\rho\sigma}
          +\tfrac18(\beta+2)(3\beta-2)\,\Rho^2
          \big]\, .
\end{eqnarray*}
Even for a Ricci-symmetric space (where the derivative terms drop),  no choice of~$\beta$ removes all remaining terms quadratic in the Schouten tensor and its trace.
[We  see no strong physical motivation to single out backgrounds with covariantly constant Einstein tensor  subject to a further quadratic curvature constraint.] This last detour reassures us that no interesting, at best slightly more general than Einstein, backgrounds are allowed.  

\section{PM self-interaction?}
 
We emphasize at the outset that the aim of this section is to study putative self-interacting extensions of~PM solely within the context of the~CG framework. That is, our results---which
will  face the usual stringent limitations on such extensions---strictly apply only to this framework,
although they are suggestive, and the allowed nonlinearities are quite efficiently generated. 
We will need a version of the~CG action that is more useful for our purposes, in which the~PM field is clearly isolated. This is accomplished by~CG's ``Ostrogradsky'' second order formulation~\cite{Kaku:1977pa},
\begin{equation*}\label{kaku}
S[g,\varphi]=-\int\sqrt{-g}\,
\Big[ \tfrac{\Lambda}6\left(R-2\Lambda\right)
+\varphi^{\mu\nu}\,\G_{\mu\nu}
+\varphi^{\mu\nu}\,\varphi_{\mu\nu}-\varphi^2\Big]\,,
\end{equation*}
where~$\G_{\mu\nu}:= G_{\mu\nu}+\Lambda\,g_{\mu\nu}$ is 
the cosmological Einstein tensor. 
Upon completing the square, we see that  the auxiliary field becomes the cosmological Schouten tensor~\eqn{varphi}.
To analyze the spectrum of the theory about an Einstein background~$\bar g_{\mu\nu}$ with cosmological constant~$\Lambda$, we  
linearize  in metric perturbations~$h_{\mu\nu}=g_{\mu\nu}-\bar g_{\mu\nu}$. 
Keeping terms quadratic in fluctuations and making the field redefinition
\begin{equation}
h_{\mu\nu}\,\to\,h_{\mu\nu}+\tfrac6\Lambda\,\varphi_{\mu\nu}\,,
\label{lin_redef}\end{equation}
yields the action (the metrics appearing in ${\bm G}$ and ${\bm F}$  are set to $\bar g_{\mu\nu}$)
\begin{eqnarray}
	S^{\sss\rm (2)}[h,\varphi]
	=-\tfrac14\int\!\!\sqrt{-\bar g}\,  
	\Big[\tfrac\Lambda6\,h^{\mu\nu}\,{\bm G}_{\mu\nu}^{\rho\sigma}h_{\rho\sigma}
	-\tfrac{6}\Lambda\,\varphi^{\mu\nu}
	\left({\bm G}_{\mu\nu}^{\rho\sigma}
	-\tfrac23\,{\Lambda}\,{\bm F}^{\rho\sigma}_{\mu\nu}\right)\varphi_{\rho\sigma}\Big]\, . \label{lin_act}
\end{eqnarray}
Here~$-{\bm G}_{\mu\nu}^{\rho\sigma}\,h_{\rho\sigma}/2$ is the linearized 
cosmological Einstein tensor defined in~\eqn{Gop} and all indices are moved by~$\bar g_{\mu\nu}$.
The Pauli--Fierz (PF) mass operator is defined as
${\bm F}^{\rho\sigma}_{\mu\nu}:=\delta^\rho_{\mu}\,\delta^\sigma_{\nu}-
g_{\mu\nu}\, g^{\rho\sigma}$, so the~PM field equation is
$$\big({\bm G}_{\mu\nu}^{\rho\sigma}-\tfrac23\,{\Lambda}\,{\bm F}^{\rho\sigma}_{\mu\nu}\big)\,\varphi_{\rho\sigma}=0\, .$$ 
Thus, the first term of~\eqn{lin_act} is linearized Einstein--Hilbert,
while the terms with round brackets (the sum of the linearized gravity kinetic term and a  Pauli--Fierz mass term tuned
to the~PM value~$m^2=2\Lambda/3$) give the~PM theory, all in an Einstein background. 
Hence the model describes the ``difference'' of  massless and~PM excitations. 
Moreover, integrating out (at linear level) the field~$\varphi_{\mu\nu}$ appearing before the field redefinition~(\ref{lin_redef}), gives the fourth order equation
$${\bm B}_{\mu\nu}^{\rho\sigma}\,h_{\rho\sigma}=0\, ,\qquad\mbox{where}\qquad {\bm B}_{\mu\nu}^{\rho\sigma}:={\bm G}_{\mu\nu}^{\alpha\beta}\, {\bm F}^{-1}{}_{\alpha\beta}^{\gamma\delta}\, {\bm G}^{\rho\sigma}_{\gamma\delta}
-\tfrac23\,{\Lambda}\, {\bm G}_{\mu\nu}^{\rho\sigma}\, ,$$ 
for the original metric fluctuations. Indeed,~${\bm B}_{\mu\nu}^{\rho\sigma}\,h_{\rho\sigma}$
is the Bach tensor linearized about an Einstein background.

The relative sign
of the two parts of the linearized action~\eqn{lin_act} reflects the unavoidable relative ghost  structure. In particular, states with~$\varphi_{\mu\nu}=0$
constitute a unitary, massless spin~$s=2$ spectrum. When the cosmological constant is positive (dS), states with~$h_{\mu\nu}=0$ correspond to a unitary
PM  spectrum. We now proceed to study the latter truncation;
a key step is to understand the model's gauge structure. At linear level, the graviton~$h_{\mu\nu}$ enjoys a linearized diffeomorphism
symmetry\footnote{As an aside, we observe that the derivation of the linear~PM model from Weyl invariant~CG theory gives a novel proof of the $SO(4,2)$ conformal invariance of~PM excitations. (In fact, conformal invariance was the original rationale behind the~PM model~\cite{Deser:1983tm}, and is enjoyed by all maximal depth, four-dimensional~PM theories of generic spin~\cite{Deser:2004ji}.) In detail, whenever a field is coupled to the metric, maintaining Weyl invariance, then setting the metric to a background yields an action that enjoys any conformal isometries as symmetries. Thus the non-linear model generated by setting the metric in~\eqn{kaku} to a background is guaranteed to enjoy this  symmetry; since it holds order by order in~$\varphi$, it is also a symmetry of  linearized~PM. }~$\delta h_{\mu\nu}=\nabla_{\mu}\,\xi_{\nu}+\nabla_{\nu}\,\xi_{\mu}$ 
while the~PM field~$\varphi_{\mu\nu}$ transforms according to the double derivative scalar variation~\eqn{pm_gauge}; at linear level each field is inert under the other's transformations.
In fact, the~PM gauge symmetry is inherited from the Weyl symmetry of~CG.
The full non-linear action~\eqn{kaku} is invariant under both gauge transformations,
\bea
	\delta g_{\mu\nu}&=&\nabla_{\mu}\,\xi_{\nu}+\nabla_{\nu}\,\xi_{\mu}+2\,\alpha\,g_{\mu\nu}\,,\nn\\[2mm]
		\delta \varphi_{\mu\nu}&=&
	{\mathcal L}_{\xi}\,\varphi_{\mu\nu}
	+\big(\nabla_{\mu}\,\partial_{\nu}+\tfrac\Lambda3\,g_{\mu\nu}\big)\,\alpha\, .\label{nonlin tr}
\eea	
The metric transformation is now a sum of diffeomorphism  and  Weyl transformations as is the~$\varphi_{\mu\nu}$ transformation:
${\mathcal L_{\xi}}$ is the Lie derivative along the vector field~$\xi$ and the Weyl term follows from the transformation of the Schouten tensor~\eqn{rhoweyl}. 

Without incurring the ghost problem of~CG, 
 we may search for some combination of fields that, when held to an appropriate background, yields a consistent
truncation to a self-interacting~PM model.\footnote{Indeed, the converse version of this procedure can be applied to 
produce cosmological gravity from
CG for the full, non-linear theory:
Examining the gauge transformations~\eqn{nonlin tr}, we see that the~PM background~$\varphi_{\mu\nu}=0$
is preserved by diffeomorphisms but not Weyl transformations. Hence, setting~$\varphi_{\mu\nu}=0$ yields a diffeomorphism 
invariant theory; performing this substitution in the action~\eqn{kaku} gives cosmological Einstein gravity.}
We must now find the proper  combination of fields to set to a background that yields the desired decoupling.
At linear level, the answer to this requirement is given by the field redefinition~\eqn{lin_redef}. There,  the choice for the metric
fluctuations~$h_{\mu\nu}=0$ is respected by~PM gauge transformations. This substitution in the linearized action~\eqn{lin_act}
yields the free~PM action in an Einstein background.
Therefore we begin by positing a candidate for a
 non-linear version of the field redefinition~\eqn{lin_redef} (that mixes~$g_{\mu\nu}$ and~$\varphi_{\mu\nu}$) such that a consistent~PM theory results from holding the redefined metric
to a suitable fixed value:
\begin{equation}\label{field_redef}
\left\{
\begin{array}{ccl}
g_{\mu\nu}&\to&g_{\mu\nu}+\tfrac6\Lambda\,\varphi_{\mu\nu}\\[2mm]
\varphi_{\mu\nu}&\to& \varphi_{\mu\nu}\,.
\end{array}\right.
\end{equation}
[We could have allowed for further redefinitions of both fields, by adding (to each) initially arbitrary functions starting at second order, so as to preserve the linear choice~\eqn{lin_redef}, but in fact this would only affect quartic corrections, and we will, for good reason,  stop at cubic order.] 
With this field redefinition, the~CG action~\eqn{kaku} reduces to that of a 
``matter'' field~$\varphi_{\mu\nu}$ coupled to a (dynamical) metric:
$$
	S[g,\varphi]=\int \sqrt{-g}\,\big[
	-\tfrac{\Lambda}6\,(R-2\,\Lambda)
	+\tfrac{6}\Lambda\,\mathscr L_{\rm\sss~PM}(\varphi,\nabla\varphi)\,\big]\,,
$$
where~$\mathscr L_{\rm\sss~PM}$ is the candidate~PM Lagrangian.
Its $\varphi_{\mu\nu}$ dependence  
is highly non-linear, with self-interactions coming from re-expressing all the original metric dependence of the action~\eqn{kaku}
in terms of the shifted combination~$g_{\mu\nu}+\frac6\Lambda\, \varphi_{\mu\nu}$. After making this expansion, we set~$g_{\mu\nu}$ to 
any Einstein metric with cosmological constant~$\Lambda$. This leaves us with the~PM candidate
\begin{equation*}
S_{\rm \sss~PM}[\varphi]=\tfrac{6}\Lambda\,
\int \sqrt{-g}\,\mathscr L_{\rm\sss~PM}(\varphi,\nabla\varphi)\, ,
\end{equation*}
to be computed as an expansion in
$\varphi_{\mu\nu}$:
\begin{equation*}
	\mathscr L_{\rm\sss~PM}
	=\tfrac14\,\varphi^{\mu\nu}\left({\bm G}_{\mu\nu}^{\rho\sigma}
	-\tfrac23\,{\Lambda}\,{\bm F}^{\rho\sigma}_{\mu\nu}\right)\varphi_{\rho\sigma}
	+	\sum_{n=3}^{\infty}\mathscr L_{\rm\sss~PM}^{\sss(n)}\,.
\end{equation*}
The absence of a term linear in~$\varphi_{\mu\nu}$  follows from the linearized analysis and relies on the fact that~$g_{\mu\nu}$ is now an Einstein metric. 

Before presenting our explicit cubic vertices, let us show that there is no fully non-linear truncation of~CG to an interacting 
PM theory.
(This neither annuls consistency of the cubic vertices with respect to linearized gauge 
transformations, nor rules out any other  ultimate theory of self-interacting~PM fields.) To determine whether a truncation that  
takes~$g_{\mu\nu}$ to be a fixed Einstein background is consistent, we must study the gauge invariances of the theory.
The precise form of the 
 underlying~CG gauge transformations in terms of the redefined fields~(\ref{field_redef}) is:
\begin{eqnarray*}\label{nonlin tr 2}
	&&
	\delta g_{\mu\nu}=\ {\cal L}_\xi g_{\mu\nu}\, 
	-\, \tfrac6\Lambda\,\big[\,\nabla_{\mu}\,\partial_{\nu}
	+\tfrac6\Lambda\,[(g+\tfrac6\Lambda\,\varphi)^{-1}]^{\rho\sigma}\,
	\gamma_{\rho\mu\nu}\,\partial_{\sigma}\,\big]\,\alpha\,,\\[2mm]
	&&
	\delta \varphi_{\mu\nu}=
	\mathcal L_{\xi}\,\varphi_{\mu\nu}
	+\big[\,\nabla_{\mu}\,\partial_{\nu}+\tfrac\Lambda3\,g_{\mu\nu}
	+\tfrac6\Lambda\,[(g+\tfrac6\Lambda\,\varphi)^{-1}]^{\rho\sigma}\,
	\gamma_{\rho\mu\nu}\,\partial_{\sigma}
	+2\,\varphi_{\mu\nu}\,\big]\,\alpha\, .\nonumber
\end{eqnarray*}
Here we have denoted the Christoffel symbols
of~$\varphi_{\mu\nu}$, covariantized with respect to~$g_{\mu\nu}$,~by
\begin{equation*}
	\gamma_{\rho\mu\nu}:=
	\tfrac12\left(\nabla_{\mu}\varphi_{\nu\rho}+\nabla_{\nu}\varphi_{\mu\rho}
	-\nabla_{\rho}\varphi_{\mu\nu}\right)\,.
\end{equation*}
Firstly observe  that at  leading order in~$\varphi$, the choice of diffeomorphism parameter~$\xi_\mu = 3\,\partial_\mu \alpha/\Lambda$ cancels the Lie derivative term
${\cal L}_\xi g_{\mu\nu} = \nabla_{\mu}\,\xi_{\nu}+\nabla_{\nu}\,\xi_{\mu}$ against the double gradient of the scalar parameter~$\alpha$ in the metric variation.
This is just a restatement of our linear result that the dynamical metric can be decoupled (at that order), leaving  linear~PM. Consistency of the
non-linear truncation requires that there exist a choice of~$\xi$ achieving this cancellation to all orders. This would determine the higher order terms in the variation of~$\varphi$, leaving the~PM action~$S_{\sss\rm~PM}[\varphi]$ invariant. To establish a no-go result, we need only show that already 
no choice of~$\xi$ achieves this cancellation for the next-to-leading order terms in~$\varphi$ in the metric variation.
Focussing on the~$\gamma^{\rho}_{\mu\nu}\partial_{\rho}\alpha$ part of~$\delta g_{\mu\nu}$ that is linear in~$\varphi$ , we immediately see that it
can never be written as~$\nabla_{(\mu}X_{\nu)}$, for any~$X_{\nu}$ even on~PM-shell.
This establishes our claimed no-go result for truncating~CG to a~PM theory beyond linear order.

Finally, we compute the cubic  vertices, which, being  guaranteed  invariant under  leading~PM gauge transformations 
$\delta \varphi_{\mu\nu} = \big(\nabla_\mu\partial_\nu+\frac\Lambda 3g_{\mu\nu}\big) \alpha$,  are  candidate vertices for a putative non-linear self-inter\-acting~PM theory.
The form of~$n$-th order Lagrangian of the~PM field determined by the field redefinition~\eqn{field_redef} can be
obtained from the following correspondence,
\begin{eqnarray}\label{n-th vertex}
	 (\tfrac\Lambda6)^{n+1}\sqrt{-g}\ \mathscr L^{\sss (n+2)}_{\rm\sss~PM}
	&=&\tfrac{n+1}{(n+2)!}\,\varphi_{\mu\nu}\, \delta^{n+1}_{g|_\varphi}\!
	\left[\sqrt{-g}\, \cal G^{\mu\nu} \right]
	 \\[2mm]
	&&+\,
	\tfrac\Lambda6\,\tfrac1{n!}\delta^{n}_{g|_\varphi}\left[\sqrt{-g} \, g^{\mu\nu}g^{\rho\sigma}\right]\ 
        \left(\varphi_{\mu\rho}\,\varphi_{\nu\sigma}-
	\varphi_{\mu\nu}\,\varphi_{\rho\sigma}\right).\nn
\end{eqnarray}
Here~$\delta^{n}_{g|_\varphi}$ signifies taking the~$n$-th variation with respect to the  metric and then replacing~$\delta g_{\mu\nu}$ by~$\varphi_{\mu\nu}$; the result is 
of order $n$ in~$\varphi_{\mu\nu}$. 
In the first line, we have used the fact that the first metric variation of the cosmological Einstein--Hilbert action produces the cosmological Einstein tensor~${\cal G}_{\mu\nu}$,
which allows~$(n+2)$ variations of that term to be combined with~$(n+1)$ variations of the coupling of the cosmological Einstein tensor to the~PM field in~\eqn{kaku}.
If we evaluate the above interaction Lagrangians explicitly then, since
they are given in terms of multiple variations of the Ricci tensor,  the generic outcome for~$\mathscr L_{\sss\rm~PM}$ is 
a two-derivative self-coupling of~$\varphi_{\mu\nu}$, a curvature coupling
and a potential for~$\varphi_{\mu\nu}$\,.
We also note that multiplying the original~CG action~\eqn{W2action} by the dimension-free combination $\Lambda^{-1}\kappa^{-2}$ of the cosmological and gravitational constants,
and redefining the~PM field $\varphi\to\Lambda\,\kappa\,  \varphi$ gives, schematically, the canonically normalized action $$S\sim \frac1{\kappa^2} \int (R-2\Lambda) +\int \Big[(\nabla \varphi)^2+\Lambda \, \varphi^2 \Big]+ \sum_{n=3}^\infty \kappa^{n-2} \big[\varphi^{n-2}\,\nabla \varphi\, \nabla \varphi  + \Lambda\,  \varphi^{n}\big]\, .$$

Now, let us focus on computing the cubic part~$\mathscr L_{\rm \sss~PM}^{\sss (3)}$ in~(\ref{n-th vertex}).
Note that  since we work on an Einstein background, we may set  ${\cal G}_{\mu\nu}=0$ (when it is not varied); also, since we only quote the vertex up to a possible
field redefinition, at this order we may use the linear~PM field equation, 
which can be written as\footnote{Notice that the cubic vertex, therefore,  schematically takes the form $$S^{\sss(3)}_{\sss\rm~PM}=\delta_{g|_\varphi} S^{\sss(2)}_{\sss\rm~PM}\,  +\, \int  \varphi^3\, ,$$ where $S^{\sss(2)}_{\sss\rm~PM}$ is the leading order~PM action and $\varphi^3$ 
denotes cubic potential terms in $\varphi_{\mu\nu}$.}  
$
\delta_{g|_\varphi} {\mathcal G}_{\mu\nu}+\tfrac{\Lambda}{3} \big(\varphi_{\mu\nu} - g_{\mu\nu}\varphi\big)=0\,
$.
Moreover, since the vertex is cubic in $\varphi_{\mu\nu}$, we may write 
$$
\tfrac6\Lambda\,  T^{\mu\nu}:= \frac13\,  \frac{1}{\sqrt{-g}} 
\frac{\delta S^{\sss (3)}_{\sss\rm~PM}}{\delta \varphi_{\mu\nu}}\, ,\qquad
S^{\sss (3)}_{\sss\rm~PM}=\tfrac 6\Lambda\int \sqrt{-g} \ \varphi_{\mu\nu}\,  T^{\mu\nu} \, .
$$
By construction, $S^{\sss(3)}_{\sss\rm~PM}$ is invariant under the linear order~PM gauge transformation~\eqn{pm_gauge} modulo the linear field equations. This guarantees that $T_{\mu\nu}$ obeys the Noether identity
\be\label{noether}
(\nabla^\mu\nabla^\nu +\tfrac \Lambda3\,g^{\mu\nu})\,
T_{\mu\nu} \approx 0\, ,
\ee
in an Einstein background; here
$\approx$ denotes equality modulo the linear~PM field equations.

It remains to explicitly compute $T_{\mu\nu}$. In fact, the cubic vertex given by~\eqn{n-th vertex} at~$n=1$ is easily computed by hand.  For the Noether form of the vertex, a computer aided computation~\cite{Vermaseren:2000nd} gives
\begin{eqnarray*}
T_{\mu\nu}&\approx&
\varphi^{\rho\sigma}\, \nabla_\rho\nabla_\sigma \varphi_{\mu\nu}+\tfrac12\, \varphi_{\mu\nu}\,\Delta\varphi
-\tfrac43\, \varphi^{\rho\sigma}\,\nabla_{(\mu|}\nabla_\rho\varphi_{\sigma|\nu)}-\varphi^\rho{}_{\!\!(\mu}\,\nabla_{\nu)}\nabla_\rho\varphi \\[1mm]
&&+\,\tfrac23\,\varphi^{\rho\sigma}\,\nabla_\mu\nabla_\nu \varphi_{\rho\sigma}
+\tfrac16\,\varphi\,\nabla_\mu\nabla_\nu \varphi
+\tfrac1{6}\, g_{\mu\nu}\left( \varphi^{\rho\sigma}\,\nabla_\rho\nabla_\sigma \varphi-\varphi\, \Delta\varphi\right)\\[1mm]
&&+\, \nabla^\rho \varphi\, (\tfrac 32\,  \nabla_\rho \varphi_{\mu\nu} -\tfrac 23 \, \nabla_{(\mu}\varphi_{\nu)\rho})
-\tfrac13\, \nabla^\rho \varphi^\sigma{}_{\!\!\mu}\, \nabla_\rho \varphi_{\sigma\nu}
- \nabla^\rho \varphi^\sigma{}_{\!\!(\mu|} \,\nabla_\sigma \varphi_{|\nu)\rho} \\[1mm]
&&+\,\tfrac23\, \nabla_{(\mu|}\varphi^{\rho\sigma}\, \nabla_{\rho|}\varphi_{\nu)\sigma}
+\tfrac16\, \nabla_{\mu}\varphi^{\rho\sigma}\, \nabla_{\nu}\varphi_{\rho\sigma}-\tfrac13 \, \nabla_\mu\varphi\, \nabla_\nu \varphi \\[1mm]
&&-\, g_{\mu\nu}\,(\tfrac5{12}\, \nabla^\rho\varphi^{\sigma\tau}\,\nabla_\rho\varphi_{\sigma\tau}
-\tfrac12\,  \nabla^\rho\varphi^{\sigma\tau}\,\nabla_\sigma\varphi_{\rho\tau}+\tfrac1{12}\, \nabla^\rho \varphi\, \nabla_\rho \varphi) \\[1mm]
&&-\, \Lambda \, (\tfrac1 {18}\, \varphi \, \varphi_{\mu\nu} + \tfrac{5}9\, \varphi^\rho{}_{\!\! \mu}\, \varphi_{\nu\rho})
+\Lambda\,  g_{\mu\nu}\, ( \tfrac{11}{36}\, \varphi^{\rho\sigma}\,\varphi_{\rho\sigma}-\tfrac1{36} \, \varphi^2 ) \\[1mm]
&&-\,\tfrac23 \, W^{\rho\tau}{}_{(\mu}{}^\sigma \varphi_{\nu)\tau}\,\varphi_{\rho\sigma}
-\tfrac23 \, W^\rho{}_{(\mu\nu)}{}^\sigma \varphi^\tau{}_{\!\! \rho}\,  \varphi_{\tau\sigma} 
-\tfrac13\, g_{\mu\nu}\, W^{\rho\tau\kappa\sigma}\varphi_{\rho\sigma}\,\varphi_{\tau\kappa}\, .
\end{eqnarray*}
As a check, we verified that this $T_{\mu\nu}$ obeys the Noether identity~\eqn{noether} for
constant curvature backgrounds (vanishing Weyl tensor).

As stated at the start of this Section, our cubic results were obtained entirely within the~CG framework. However, their consistency is independent of their origin, since they are of course disjoint from any higher-order problems. Indeed, the vertex $S^{(3)}_{\sss\rm~PM}$ was constructed by a St\"uckelberg method
 in~\cite{Zinoviev:2006im}, where it was also shown that two-derivative~PM self-interactions  exist only for $d=4$, which dovetails perfectly with their~CG origin uncovered here. These results  also fit with the recent
 work of~\cite{Joung:2012rv} where all consistent cubic interactions 
 (not necessarily  two-derivative ones)
 involving~PM fields of generic spin were considered. There 
  it was shown that for generic dimensions there are only two~PM self-couplings
  involving at most four and six derivatives respectively.
 However, precisely in four dimensions,  the Gau\ss-Bonnet identity reduces the maximal
 four-derivative coupling to a two-derivative one.\footnote{In fact, for constant curvature backgrounds, the Cotton-like tensor~\cite{Deser:2006zx} 
 $$F_{\mu\nu}{}^\rho:=\nabla_{\mu} \varphi_{\nu}{}^{\!\rho} - 
 \nabla_{\nu}\varphi_\mu{}^{\!\rho}$$ is invariant under~PM gauge transformations~\eqn{pm_gauge}. (Strictly this version of the Cotton tensor is not the metric one, because the~PM field is not the Schouten tensor, although in the underlying~CG setting this 
is  in fact the case.)
 Therefore any quartic derivative order, cubic vertex of type $\int (\nabla F) F F$ is~PM invariant. In four dimensions, it should be possible to employ the Gau\ss--Bonnet identity to write this as a manifestly invariant cubic vertex quadratic in derivatives.}

\section{Conclusions}

We have used $d=4$ conformal Weyl gravity as a tool to explore the extent of the usual higher spin constraints on~PM self-and gravitational- couplings. 
We concluded that these obstructions were indeed present here as well:  first, no backgrounds more general than Einstein 
were permitted for~PM's propagation.
Then, we exploited the truncation of~CG to~PM in a fixed geometry to find what ghost-free self-couplings, if any, might be permitted within the~CG framework. Although relative ghost-like graviton modes could be removed at linear order leaving (consistent) linear~PM, in contrast to the~PM truncation of~CG to cosmological gravity~\cite{Maldacena:2011mk}, the gauge structure of~CG does not allow the graviton truncation to continue to higher orders.
An old problem (one that already occurs in similar attempts at extending  other higher-spin) has struck again:  despite the  possibility of a lowest order invariant cubic self-interaction (expressed as the coupling of the quadratic Noether current maintaining the initial Abelian invariance to the field amplitude), self-coupling inconsistencies set in at quartic order. 
CG underlies cosmological Einstein gravity  but  it does not truncate to a non-linear ``PM general relativity''.  
 Despite the results achieved here, we should emphasize that they merely begin to reflect
CG's potential to explore (A)dS models' physical content in a direct way. The underlying 
CG~technology is clearly capable of yielding far more insight.

No-go theorems are notorious for their loopholes. Spin (2,3/2)-gravity
and supergravity theories circumvent just such higher-spin pitfalls~\cite{Buchdahl} while for (towers of) massive higher spins, string theory provides presumably consistent interactions; infinite 
towers of massless higher spins  can also be written in (A)dS backgrounds~\cite{Vasiliev}. 
Nonetheless,  our results relying on~CG as the underpinning of~PM self-interactions seem quite 
robust; they  agree with the claim of~\cite{Zinoviev:2006im} that
 it is impossible to proceed beyond cubic order for the two-derivative PM theory.

One interesting feature of~CG is that the~PM field can be consistently turned off, leaving cosmological Einstein gravity (at least classically). 
In other words, without additional matter couplings, 
choosing  initial  conditions such that~$\varphi_{\mu\nu}$ is zero at some initial time, 
it will remain  trivial  while the metric~$g_{\mu\nu}$ can realize 
any  Einstein solution~\cite{Maldacena:2011mk}. This suggests the converse truncation:  a situation where the~PM field~$\varphi_{\mu\nu}$ is not strictly zero but rather  nearly zero in
some arbitrarily large time interval~$t_{i}\ll t_{f}$.
Cosmology would then have approximate Einstein behavior for that epoch, while in the region~$t  \ll t_{i}$ or~$t \gg t_{f}$, non-Einstein solutions could emerge.
(The consequences for cosmological expansion with a partially conserved symmetric two index boundary operator were also considered in~\cite{Dolan:2001ih}.)~CG~could then be used to generate transitions from a~dS inflationary behavior of the cosmic scale factor to  one controlled by~PM modes.
Ghosts and loss of stability at early and late times may even be a useful/acceptable feature in this scenario.

A separate speculation is that gravity-like, or even self-interacting~PM-like models for higher~$s>2$, might be achievable by studying higher-spin versions of~CG. Indeed, interacting conformally invariant higher-spin models that can be viewed as analogs of~CG do exist~\cite{Segal:2002gd, Bekaert:2010ky}. Perhaps a higher spin version of our approach could could be fruitfully applied to them.

\section*{Acknowledgements}
We thank Hamid Afshar, Rod Gover, Daniel Grumiller and Karapet Mkrtch\-yan for fruitful discussions.
EJ and AW acknowledge the ESI Vienna Workshop on Higher Spin Gravity. 
SD was supported in part by NSF PHY-1064302 and DOE DE-FG02-164 92ER40701 grants.
The work of EJ was supported in part by Scuola Normale Superiore, by INFN (I.S. TV12) and by the MIUR-PRIN contract 2009-KHZKRX.

\end{document}